# Ultrafast optical rotations of electron spins in quantum dots

A. Greilich[1*], Sophia E. Economou[2], S. Spatzek[1], D. R. Yakovlev[1,3], D. Reuter[4], A. D. Wieck[4], T. L. Reinecke[2], and M. Bayer[1]

*[1] Experimentelle Physik 2, Technische Universität Dortmund, 44221 Dortmund, Germany*
*[2] Naval Research Laboratory, Washington, D.C. 20375, USA*
*[3] A. F. Ioffe Physico-Technical Institute, Russian Academy of Sciences, 194021 St. Petersburg, Russia*
*[4] Angewandte Festkörperphysik, Ruhr-Universität Bochum, 44780 Bochum, Germany*

*[*] e-mail: alex.greilich@udo.edu*

**Coherent manipulation of quantum bits (qubits) on time scales much shorter than the coherence time[1,2] is a key prerequisite for quantum information processing. Electron spins in quantum dots (QDs) are particularly attractive for implementations of qubits. Efficient optical methods for initialization and readout of spins have been developed in recent years[3,4]. Spin coherence times in the microsecond range have been demonstrated[5], so that spin control by picosecond optical pulses would be highly desirable. Then a large number of spin rotations could be performed while coherence is maintained. A major remaining challenge is demonstration of such rotations with high fidelity. Here we use an ensemble of QD electron spins focused into a small number of precession modes about a magnetic field by periodic optical pumping. We demonstrate ultrafast optical rotations of spins about arbitrary axes on a picosecond time scale using laser pulses as control fields.**

Spin rotations have been obtained for gated GaAs transport QDs using rf fields, but such manipulations are slow with times comparable to the spin coherence time[6]. Fast optical rotation operations on spins in QDs are being sought actively, but progress to date has been limited. In an all-optical experiment on shallow interface fluctuation QDs, weak confinement precluded full unitary spin rotations[7]. Evidence for spin rotations was reported for a single GaAs interface fluctuation quantum dot using very intense optical excitation needed for spins to be rotated by precession about the effective laser magnetic field[8]. Very recently more convincing rotations in a single InAs self-organized QD were reported[9]. This work on single QDs[8,9] used pulses far from resonance, which could lead to excitation of other spins. This could inhibit the use of spins with separations less than the laser spot size for two quantum bit entanglement. In general, optical coupling to a spin in a single quantum dot is weak, and thus spin rotations on single QDs give weak responses. This generally necessitates many optical pulses in time to obtain sufficient fidelity, and it can give additional dephasing.

An ensemble of QDs has the advantage of increasing the optical coupling and can decrease the loss of information from the loss of a few spins. However, ensemble approaches typically are hampered by the inhomogeneities in QD properties, particularly in the spin splitting g-factor, and lead to dephasing of the spin precession about a magnetic field[10,11]. Recently we have demonstrated techniques for eliminating part of the inhomogeneity in the electron spin precession in ensembles of singly charged QDs[12]. We have found that at low magnetic fields the spin ensemble can be put into a Zeeman spectrum[13] that is very close to a single precession mode. This is the system that we study here. Measurements of such systems demonstrate homogeneous spin dephasing times to be at least 3 μsec (ref. 5). Thus, for optical control on the picosecond timescale, about $10^6$ operations could be carried out during the coherence time.

Our samples consist of arrays of self organized (In,Ga)As QDs each containing on average a single electron (see SI)[14]. The spins are initialized in the z direction by periodic optical laser pulses. These spins are polarized along the light propagation direction, which is parallel to QD growth direction. A magnetic field **B** is applied along x axis (Voigt geometry). Instantaneous orientation of the spins along the optical axis is obtained through optical pumping by pulses with area $\Theta = \int dt\, \mathbf{d} \cdot \mathbf{E} / \hbar$ resonant with the trion transitions (Fig. 1B). **E** is the electric field of the laser, which excites an inhomogeneous distribution of QDs with varying dipole matrix elements **d** and energies of the trion optical transition. In the present experiments the spectroscopic responses are obtained from an average over many dots. The pulse intensities are adjusted to achieve the desired average Θ in the ensemble (see Methods).

The precession of spins (Fig. 1A) is measured by *ellipticity* of a probe laser and is proportional to the spin polarization along the optical axis z. The spin vector oscillates about **B**, which is represented on the Bloch sphere by a precession in the y-z plane (Fig. 1C). We use ultrafast 'control' laser pulses to induce rotations of the spins. The axis of the spin rotation is given by the polarization of the control laser, and the rotation angle is determined by the photon energy detuning Δ of the control pulse from the optical resonance (Fig. 1D). By using $\Theta = 2\pi$ pulses and varying their detuning, we implement spin rotations by angles from $0.2\pi$ to $\pi$ and find that the detuning dependence of the angle of rotation agrees well with what we expect from theory[16,17]. By combining two spin rotations by angles of $\pi/2$ with precession-induced rotation, we obtain a composite rotation about the y axis. This can give spin rotations about arbitrary angles. For the first time for optically controlled spins, we see spin echoes and extension of the dephasing time of the spins $T^*_2$. This opens opportunities for ultrafast decoupling operations to provide increased single spin coherence times $T_2$ (refs 18, 19).

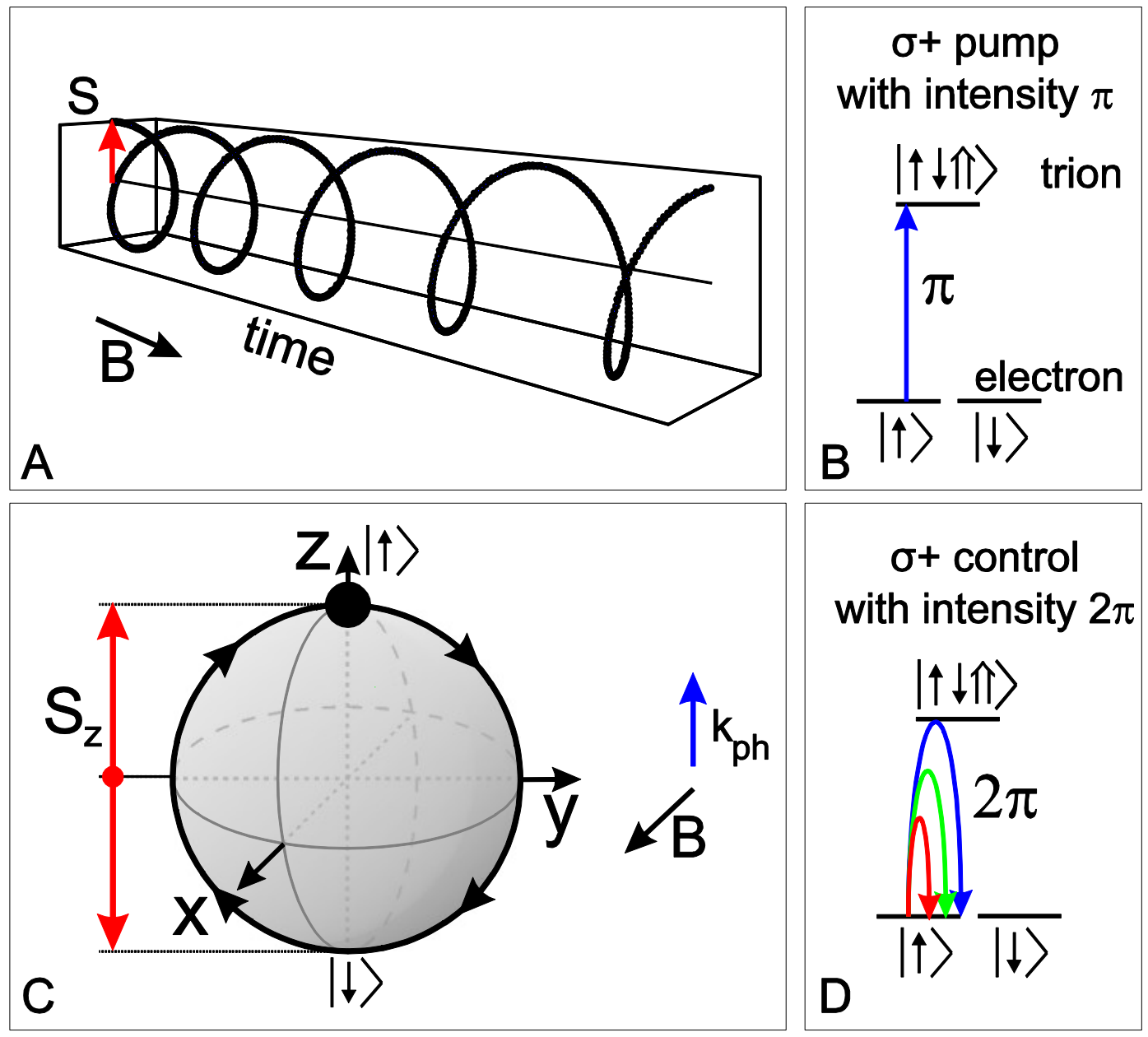

**Figure 1. Optical methods to control of spins.** (**A**) Larmor precession of a spin S about the magnetic field. (**B**) Spin polarization is generated by laser pulses resonant with the trion (two electrons and one hole) transition. Excitation with area $\Theta = \pi$ pulses is most efficient to drive the system to the trion state and to generate electron spin coherence after trion spontaneous decay[15]. (**C**) Precession of the spin around the magnetic field is represented on a Bloch sphere by the black trajectory with arrows. The measured ellipticity is proportional to the projection of electron spin polarization on the z axis ($S_z$). $k_{ph}$ is the wave vector of the pump light generating electron polarization along z axis. (**D**) A 2π control pulse does not change the population of the trion state. It provides rotation of the spins around the z axis with the angle depending on laser detuning from the trion resonance.

We used two synchronized Ti:sapphire lasers each emitting pulses of 1.5-ps duration with a separation of $T_R = 13.2$ ns. The pump and probe beams originate from the same laser (SI). The ellipticity with no control laser is shown by the bottom (black) trace in Fig. 2A vs. pump-probe delay. The precession of the spins exhibits a weak dephasing arising from the distribution of precession frequencies around the main mode $N\cdot 2\pi/T_R$ with N = 30 at B = 0.29 T (ref. 5).

The control laser is tuned in the vicinity of the trion resonance. The goal is to manipulate the spin but to avoid generating new spin polarization. Unlike the case of rf control[6], these optical pulses result in the spin state acquiring a trion component and partially exiting the qubit subspace. To return the state to the qubit subspace, the control must be unitary within that subspace. This is done by using a control pulse area of 2π, which moves the (optical) polarization through a full rotation (a Rabi flop, see Fig. 1D) to the optically excited state and back to the ground state. Then, a phase φ is induced to the ground state. Due to selection rules, this is a relative phase between the spin states pointing along +z and –z, and it is equivalent to a rotation about the z axis by angle φ (SI).

To find the pulse of area $\Theta = 2\pi$, we varied the control intensity (Fig. 2A). Figure 2B shows corresponding simulations, made by solving the complete dynamics of the three-level system of Fig. 1B (SI). Pulses with area $\Theta < \pi$ move some of the state to the excited trion level without returning it to the spin. Then the spin amplitude is reduced while its phase remains unchanged (top three curves in Fig. 2A). For $\Theta > \pi$, part of the state is returned to the spin subspace, so that the phase of the spin changes but the amplitude is reduced compared to the without control case (blue curves in Fig. 2A). The green trace in Fig. 2A is for Θ = 2π control pulse, and it has the maximum amplitude among cases with control pulses. Its amplitude is somewhat reduced from the no control case due to inhomogeneities in dipoles.

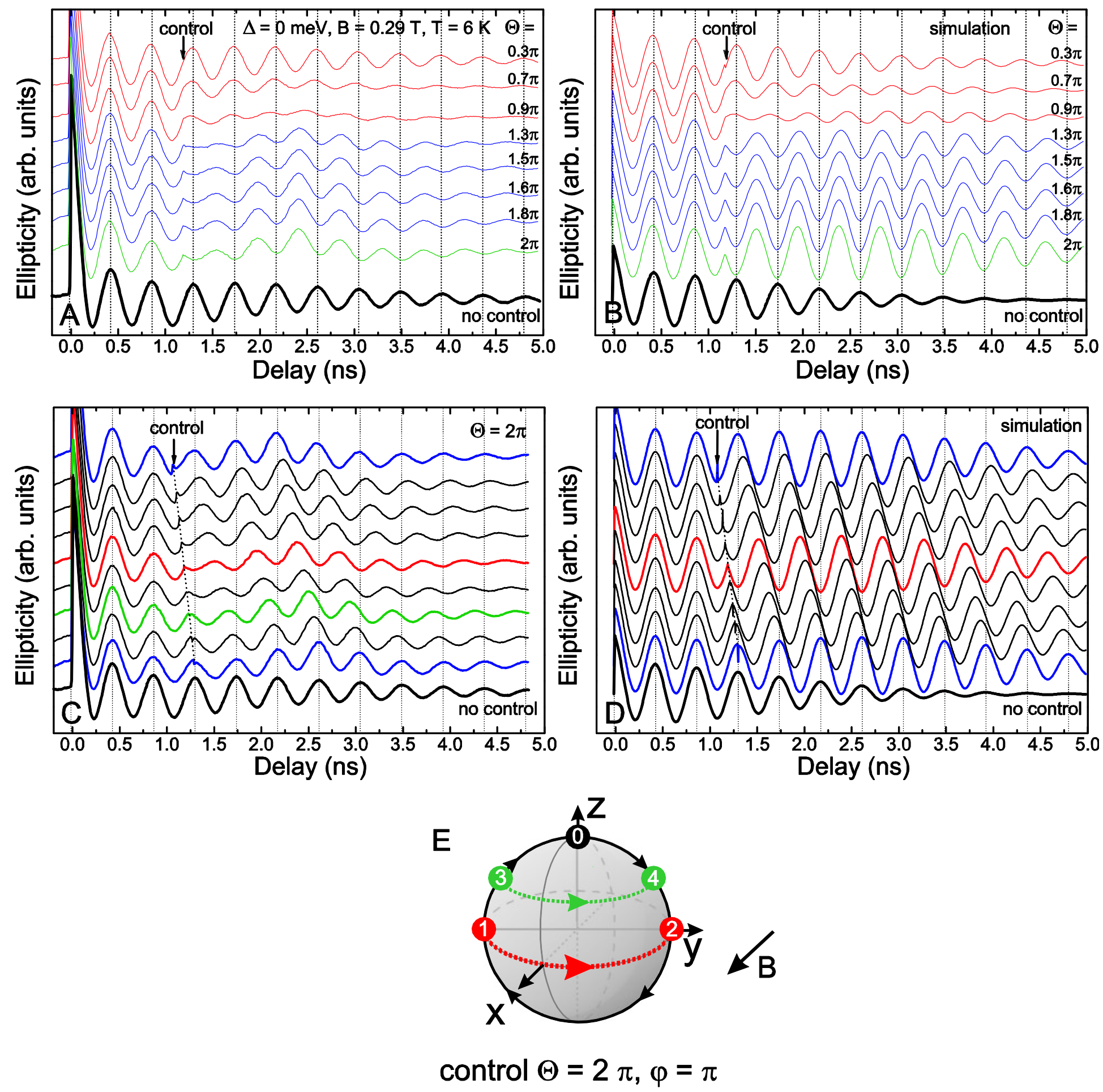

**Figure 2. Single pulse spin control.** (**A**) Measurements of ellipticity of the total spin of an ensemble vs. delay time between pump and probe. The pump and probe energies are degenerate, and the control energy is the same as that of the pump. The bottom (black) trace is without a control laser. The control hits the sample 1.2 ns after the pump, when the spin vector is along the –y axis. The control area Θ is varied to identify the 2π pulse via a phase shift of π of the spin relative to no control. (**B**) Simulations corresponding to the traces in panel A (SI). In the simulations, all QDs have the same pulse intensity. Reduction in the amplitude after the control in the experiment is attributed primarily to inhomogeneities in the dipole in different QDs, which causes part of them to see a pulse area other than Θ = 2π. The resulting error at the single qubit level could be quantified by studies of gate fidelity with quantum state tomography, which is beyond the scope of the present work. (**C**) Ellipticity measurements for control area Θ = 2π vs. pump-probe delay. The bottom (black) trace is without a control laser. Other traces are with the control at varying delays after the pump. (**D**) Simulations corresponding to panel C. (**E**) The effects of the control delay sketched on the Bloch sphere. The spin is oriented along the z axis by the pump, shown by the black dot marked with zero. It precesses about the magnetic field until it is hit by the control, which induces a π rotation about z. Then it continues its precession. The two trajectories have colors corresponding to the green and red traces of panel C.

Next we fix Θ = 2π and vary the delay τ between pump and control. This allows the spin to precess in the y-z plane for a time τ (Fig. 2C). The blue traces in Fig. 2C correspond to applying the control pulse when the spin points in the z direction, in which case it has basically no effect on the spin. For all other spin orientations the control pulse gives a π rotation about the z-axis, after which the spins continue to precess. Figure 2E shows control induced rotations for two orientations. For the red trace the control pulse was applied when the spin points in the –y direction. After the control, it is completely out of phase with the reference trace.

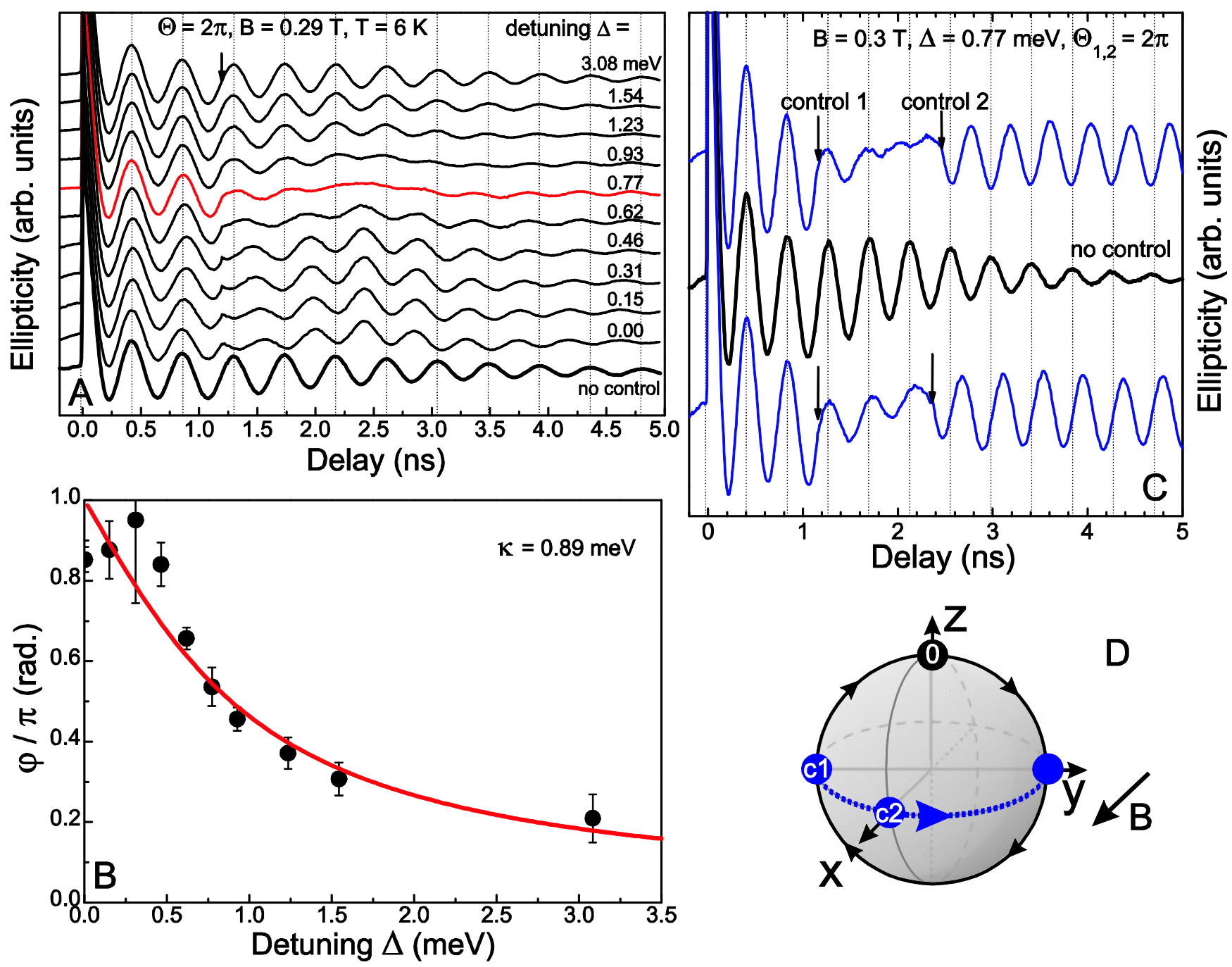

**Figure 3. Two-pulse control.** (**A**) Measurements of ellipticity for varying detunings between control and pump energies. Control pulse intensities are Θ = 2π, and the delays between the pump and control pulses were chosen so that the spin is along the –y direction when the control is applied. (**B**) Symbols give the dependence of rotation angle on the detuning between the control and pump from the data in panel A. Red line gives the theoretical result $\varphi = 2\arctan(\kappa/\Delta)$ from refs 16, 17. (**C**) Ellipticity measurements with two control pulses with varying temporal separations between them. The two control pulses have area of 2π and the same energy. Their detuning from the pump photon energy is 0.77 meV so that each of them would give a π/2 rotation about the z axis, as in the single control case. Together they have the effect of a single π rotation about the z axis. (**D**) Bloch sphere depiction of the two-control experiment. The points marked c1, c2 are where the first and second control pulses hit the spin, respectively.

Spin rotations of varying angles can be obtained by changing the detuning of the control pulse from the optical resonance while keeping its area at 2π. Figure 3A gives a series of traces with the detunings between 0 to 3.08 meV. We choose the arrival time of the control so that the net spin polarization is along –y, in the equatorial plane of the Bloch sphere. Then the effect of rotation by the control becomes most prominent in the phase and amplitude of the ellipticity signal. The angle of spin rotation φ decreases from π to nearly zero as the detuning is increased from zero to 3.08 meV (Fig. 3A). For angles of rotation $\pi/2 < \varphi < \pi$, the spin polarization immediately after the control has a component along +y and a nonzero component along the magnetic field. Then the phase of the signal is opposite to that of the reference (Fig. 3A for detunings < 0.77 meV). For $\varphi < \pi/2$, the y component is still negative right after the control, and the oscillations are in phase with the reference. The angles of rotation are obtained from the ratio of the (signed) amplitude over the amplitude of the reference at the maximum signal at 2.4 ns. It is shown as a function of detuning in Fig. 3B. This angle is in excellent agreement with the theoretical expression, which is derived analytically for pulses of hyperbolic secant temporal profile sech($\kappa \cdot t$) and gives $\varphi = 2 \arctan(\kappa/\Delta)$, where $\kappa$ is the bandwidth of the pulse[16,17].

The two experiments in blue in Fig. 3C demonstrate further the effects of spin rotations by angle $\pi/2$ obtained by using the detuning $\Delta = 0.77$ meV. For each there are two controls giving two $\pi/2$ rotations, with varying delays between them. In both the first control is applied when the spin is in the –y direction, rotating the spin into the +x direction, and the second control rotates the spin back into the y-z plane in the +y direction. In the lower case the second rotation is applied when the reference is pointing in the –z direction, which results in a phase difference of $\pi/2$ between the two applied controls and the reference (Fig. 3C). In the upper case the second control is applied when the reference spin is in the –y direction, resulting in a phase difference of $\pi$ between the two-control case and the reference.

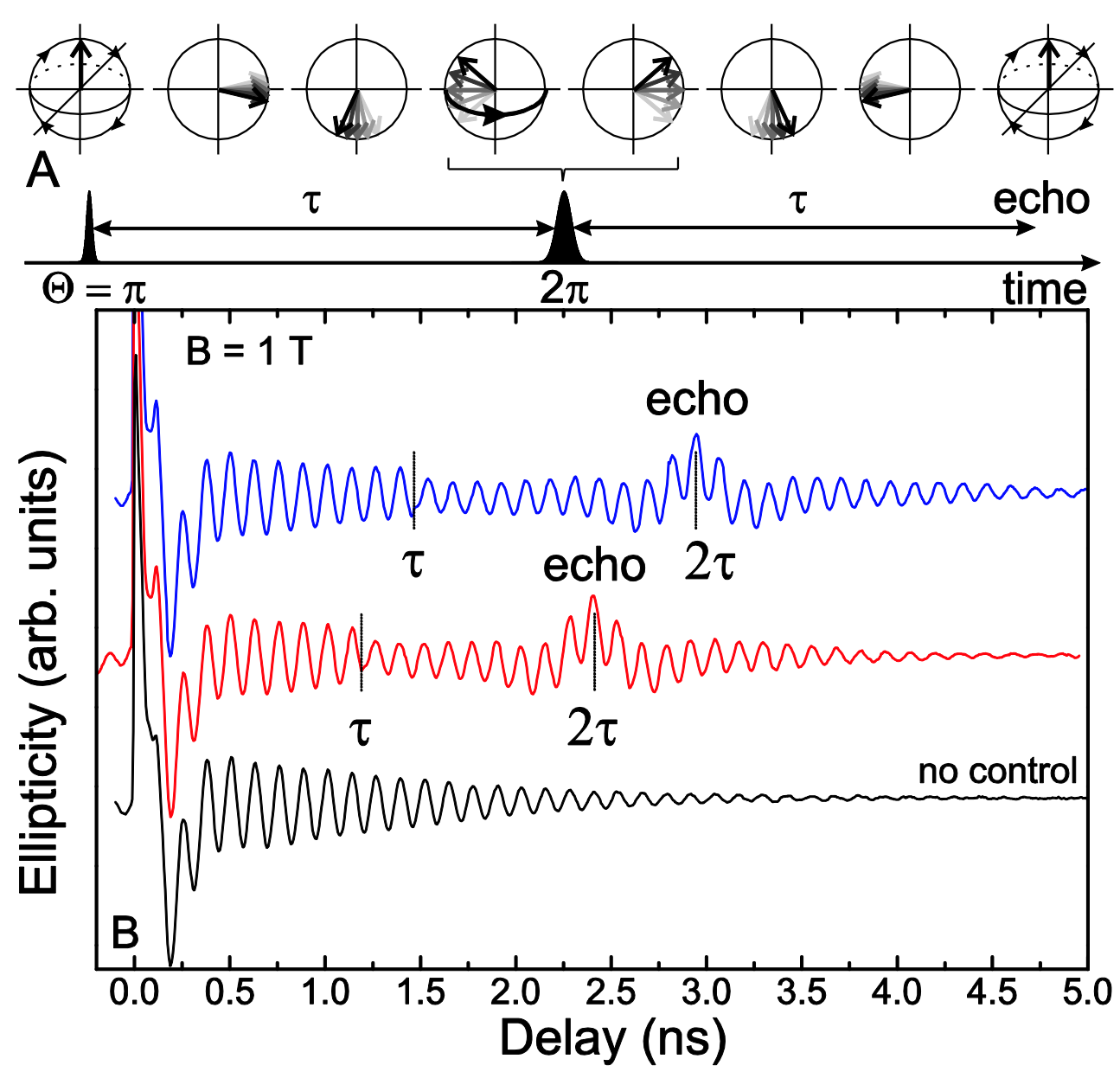

**Figure 4. Spin echoes.** (**A**) Sketch of spin echo formation vs. time in an optically controlled inhomogeneous spin ensemble. The spins are initialized along z by a pulse of area $\pi$ after which they precess about the external magnetic field with different frequencies. At time $\tau$ a control pulse of area $\Theta = 2\pi$ and zero detuning rotates the spins about the z axis by an angle $\pi$. After $2\tau$ the spins come into phase again and an echo builds up. (**B**) Experimental demonstration of a spin echo for two time delays of 1.47 and 1.20 ns between pump and control pulses.

Spin echoes are seen in Figs. 2 and 3 as an increase of signal amplitude around time delays of $2\tau$. They originate from inhomogeneities in the spin precession frequencies. A $\pi$ rotation at τ on a set of inhomogeneously precessing spins that were aligned at time zero causes the slower spins to advance ahead of the faster ones along their precession direction (Fig. 4A). Then the spins refocus at 2τ, a phenomenon known from NMR as spin echo[20,21]. The observation of electron spin echoes in our experiments demonstrates the robustness of the spin manipulations performed here. This spin echo is seen in Fig. 2A,C and also for a rotation by $\varphi > \pi/2$ (for $0 \leq \Delta \leq 0.62$ meV) in Fig. 3A. However, no echo is seen for $\varphi < \pi/2$ ($\Delta > 0.8$ meV). This is because in that case the slower spins are further delayed instead of being moved ahead of the fast ones. The echo signal became more pronounced at stronger magnetic field of 1 T, as one can see in Fig. 4B for zero detuning of the 2π control pulse and for two different delays. Thus we are able to increase the dephasing time of the ensemble, $T^*_2$, as seen by comparing the amplitude after 2τ of the rotated spins to the reference. In principle, these π rotations also can extend the $T_2$ time of the individual spins by decoupling them[18] or disentangling them[22,23] from the nuclei.

This work demonstrates robust, optically controlled single spin rotation gates on the picosecond time scale in quantum dot ensemble. These gates provide the basis for single-qubit logic operations, and thus are an important step forward in spin-based quantum dot implementations for quantum information.

The present QD experiments were done on ensembles, and they give information about the control of spins at the single dot level. The fidelity of the coherent operations could be affected by dephasing in the QDs due, e.g, to decay of optically inactive states or many-body effects, as well as to the effects of inhomogeneities in dipoles or transition energies. For applications in quantum information, these matters would need to be assessed by, e.g., full quantum state tomography, which is beyond the scope of this work.

**Methods**

Two synchronized Ti:sapphire lasers provide trains of 1.5 ps long pulses at 75.6 MHz repetition frequency, which corresponds to 13.2 ns of pulse separation. The first laser is split to give the pump and probe laser beams at the same energy. The second laser gives the additional control beams, and its energy can be tuned independently relative to the energy of pump and probe. We use a phase sensitive lock-in technique for time-resolved ellipticity measurements to trace the spin coherence excited by the circularly polarized pump. The time delay between pump and probe is obtained by a mechanical delay line. An additional delay line was used to vary the delay between the pump and control pulses.

The generation of spin coherence is most efficient for a $\Theta = \pi$ pump pulse area, while the control pulse should have area of 2π. Due to inhomogeneities of the dipole transition matrix element in the ensemble, the pulse area will not be uniform for all the dots. For the ones which see a pulse area other than 2π new spin coherence will be generated. To avoid picking up this contribution we modulate only the pump beam with a photo-elastic modulator. As can be seen from Fig. S1, this additional polarization is minimal, which allows us to use the label of $\Theta = 2\pi$ to a very good approximation when discussing the dynamics of the ensemble. Moreover, the control beam diameter was chosen to be larger than those of the pump and probe in order to achieve a high homogeneity of the control intensity over the spin coherence generated by the pump.

## Author Information

Correspondence and requests for materials should be addressed to A.G. (alex.greilich@udo.edu), S.E.E. (economou@bloch.nrl.navy.mil), T.L.R. (reinecke@nrl.navy.mil), and M.B. (manfred.bayer@physik.uni-dortmund.de).

## Acknowledgments

This work was supported by the Bundesministerium für Bildung and Forschung program "nanoquit", the Office of Naval Research, and the Deutsche Forschungsgemeinschaft. S.E.E. was an NRC/NRL Research Associate during this work.

**Supplementary Information** is linked to the online version of the paper.